\documentclass[11pt]{article}

\usepackage{setspace,graphicx,epstopdf,amsmath,amsfonts,amssymb,amsthm,versionPO}
\usepackage{marginnote,datetime,enumitem,subfigure,rotating,fancyvrb}
\usepackage{hyperref,float}
\usepackage[longnamesfirst]{natbib}
\usepackage{algorithmic}
\usepackage{algorithm}
\usdate

\excludeversion{notes}		
\includeversion{links}          

\iflinks{}{\hypersetup{draft=true}}

\ifnotes{%
\usepackage[margin=1in,paperwidth=10in,right=2.5in]{geometry}%
\usepackage[textwidth=1.4in,shadow,colorinlistoftodos]{todonotes}%
}{%
\usepackage[margin=1in]{geometry}%
\usepackage[disable]{todonotes}%
}

\makeatletter\let\chapter\@undefined\makeatother 

\usepackage{indentfirst} 
\usepackage{endnotes}    
\usepackage{jf}          
\usepackage[labelfont=bf,labelsep=period]{caption}   
\begin{document}

\setlist{noitemsep}  
\onehalfspacing      
\renewcommand{\footnote}{\endnote}  


\title{\Large \bf
	 Blockage Type Detection Process in Triangular-wave Topology for mmWave Wireless Backhaul }
 
\author{\textit{Yuchen Liu}, \textit{Qiang Hu}, and \textit{Douglas M. Blough} \\
        School of Electrical and Computer Engineering \\ Georgia Institute of Technology}

\date{}              


\maketitle





\noindent{\large \bf Abstract}
\vspace{+0.2cm}

\noindent Due to the rapid densification of small cells in 5G and beyond cellular networks, deploying wired high-bandwidth connections to every small cell base station is difficult, particularly in older metropolitan areas where infrastructure for fiber deployment is lacking.  For this reason, mmWave wireless backhaul is being considered as a cost-effective and flexible alternative that has potential to support the high-rate transmission needed to accommodate backhaul traffic demands. However, maintaining a high network survivability is extremely important since networks based on mmWave communication are susceptible to short-term blockages or node failures caused by unforeseen events. Therefore, to realize a robust mmWave backhaul network, several works propose to adopt a triangular-wave topology in roadside environments, which can eliminate the mutual interference among backhaul links, and provide robustness to blockage effects. Based on this novel topology, we investigate a blockage type detection scheme, which reacts to blockage effects quickly and assists in various path reconfiguration approaches. 

\section{Introduction}

To achieve a robust mmWave wireless backhaul with high network survivability, our work herein considers a distributed mmWave backhaul network architecture with relay nodes in urban environments. In this architecture, a number of mmWave dedicated relays in between small-cell base stations (SBSs) are deployed along urban streets, which naturally produces a mesh network structure. Note that such street-level deployment making use of lampposts has been suggested as a good choice in the 5G era [1]-[4], because it provides easy access to power, good access tier coverage for users and facilities, and ease of deployment in urban environments.

\begin{figure}[htbp]
	\vspace{-0.2cm} 
	\setlength{\abovecaptionskip}{-0.1cm} 
	\centerline{\includegraphics[width=115mm]{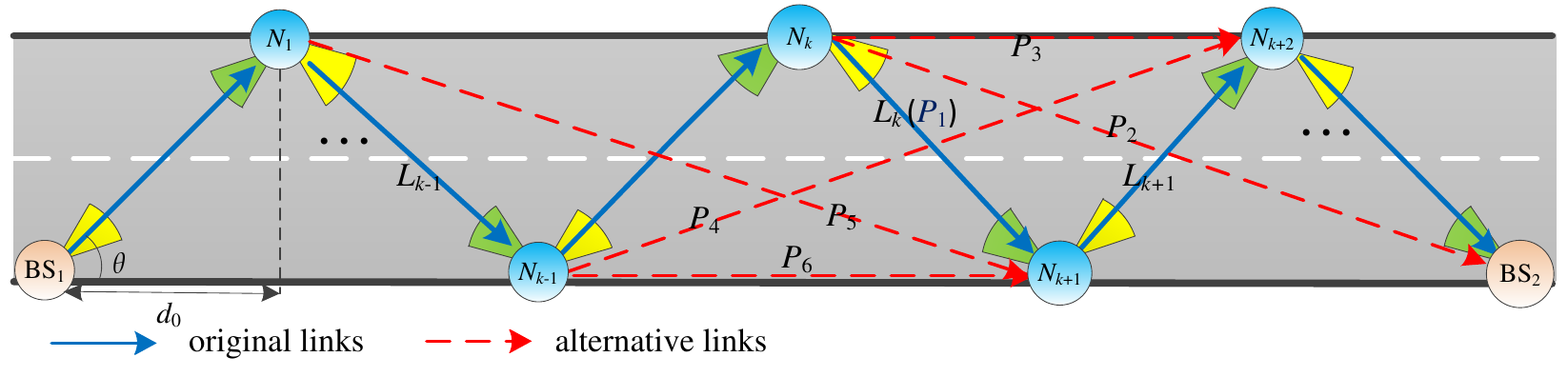}}
	\caption{Original and alternative links in the IFTW topology.}
	\label{block_topol}
\end{figure}

The interference-free triangular-wave topology (IFTW) is very well suited to provide high data rate communications for relay-assisted mmWave backhaul in roadside environments [2], [5]-[8]. In the IFTW topology, BSs and relays ($N_k$) are deployed on equally-spaced lampposts on both sides of the road, where the topology angle $\theta$ and horizontal distance between adjacent nodes $d_{0}$ are the same everywhere along the topology (as depicted by the blue links of Fig.~\ref{block_topol}). One advantage of the IFTW topology is that the mutual interference along the path can be eliminated if $\theta$ is made large enough relative to the beamwidth $\phi$ of the directional antennas (Theorem 1 in [2]), i.e., if the interference-free condition in Eq. (1) is satisfied:

\begin{equation}
\theta  - \arctan (\frac{{\tan \theta }}{3}) > \frac{\phi }{2}.
\label{interf_free_eq}
\end{equation}

Given a road width $d_{w}$ and length $d_{l}$, $d_0$ and the number of required nodes $N$ (including two BSs and several relays) in the IFTW topology are only determined by $\theta$ as ${d_0} = {d_w}/\tan \theta$ and $N = \left\lfloor {{d_l}/{d_0}} \right\rfloor  + 1$, respectively.


Another advantage of the IFTW topology is the ability to reconfigure mmWave paths to avoid obstacles that occur along the roadway.  Through adaptive beam steering when an obstacle blocks one or more of the original links, alternative links (shown in Fig.~\ref{block_topol}) can be used to restore the connectivity of the topology.  In our previous work [5], we identified the following four types of blockages that can occur from a single obstacle along the roadway with the IFTW topology:

\textit{a) Type} I: An obstacle in $L_{k}$ region blocks only a original link of the topology, such as $P_{1}$ in Fig.~\ref{block_topol}.

\textit{b) Type} II: An obstacle in $L_{k}$ region blocks an original link and an adjacent alternative diagonal link simultaneously, such as $\{P_{1}, P_{2}\}$ for $N_{k}$ or $\{P_{1}, P_{5}\}$ for $N_{k+1}$.

\textit{c) Type} III: An obstacle in $L_{k}$ region blocks an original link and a crossing alternative diagonal link, such as $\{P_{1}, P_{4}\}$.

\textit{d) Type} IV: 	An obstacle in $L_{k}$ region blocks all the TX/RX links of $N_{k}/N_{k+1}$, such as $\{P_{1}, P_{2}, P_{3}\}$ $/\{P_{1},P_{5},P_{6}\}$, which is equivalent to the failure of node $N_{k}/N_{k+1}$.

Any blockages produced by randomly placed obstacles can be decomposed into one or a combination of the above four types (Theorem 1 in [5]). This model takes into account the correlation between blockages, for example, a single blockage close to a node could block multiple mmWave paths simultaneously, which has not been considered in most previous studies [9]-[10]. In what follows, we investigate a blockage-type detection scheme based on this blockage model in the IFTW topology, which can be exploited for both link-level reconfiguration [6] and joint link-network level reconfiguration approaches [7]-[8]. 

\section{Blockage type detection scheme}

In general, the scheme has two phases: preparation phase and detection phase. The former aims to make every node in the topology know the occurrence of blockage and ready for coordination with others. The detection on link connectivity is conducted in the second phase. Since the durance of BTD and path reconfiguration is very short, we assume that no new blockage appears within this time period.

\begin{figure}[htbp]
	\vspace{-0.2cm} 
	\setlength{\abovecaptionskip}{+0.1cm} 
	\centerline{\includegraphics[width=130mm]{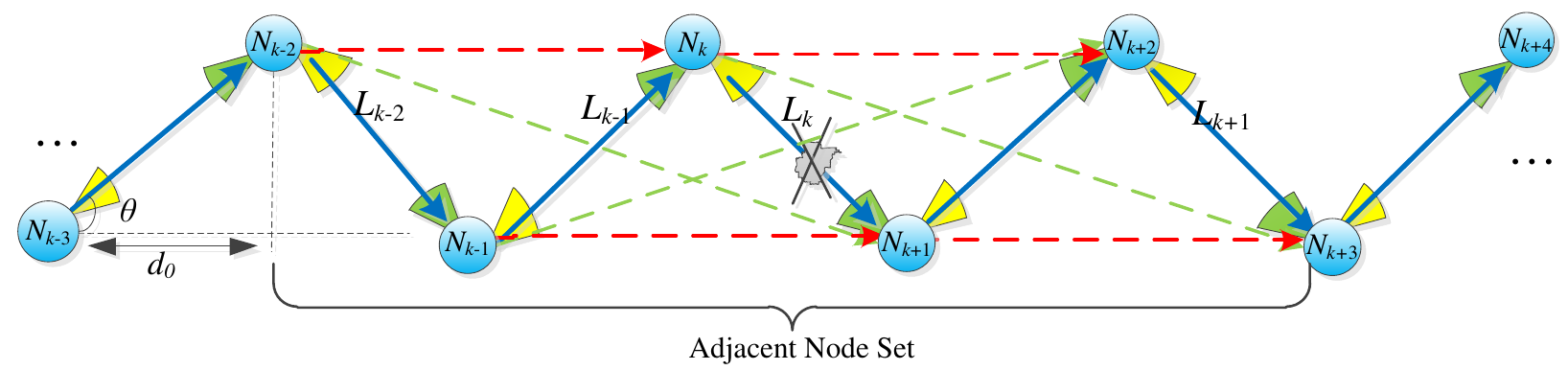}}
	\caption{The possible alternative paths and the adjacent node set of $N_{k}$.}
	\label{altern_path}
\end{figure}

%

\subsection{Preparation Phase}
For arbitrary node $N_{k}$, it knows respective adjacent node set  $\textit{Adj}$=\{$N_{k+i}$: -2$\le$$\textit{i}$$\le$3, $i$$\neq$0\}, alternative link set $\textit{Sub}$=\{$e_{k,k+j}$: 2$\le$$j$$\le$3\}, and TX link set $\textit{Tl}$=$Sub$ $\cup$ \{$e_{k,k+1}$\}. Note that \textit{Adj}, \textit{Sub} and \textit{Tl} of node $N_{k}$ may change according to different relay path reconfiguration algorithms. During this phase, the following steps describe how nodes negotiate with each other to make preparation for detection.

\textbf{\textit{Step} 1.} Considering the possible packet transmission error through links, the packet would be retransmitted for at most two times if the transmitter received nothing or NACK from receiver. Therefore, when an obstacle occurs between $N_{k}$ and $N_{k+1}$, i.e. the original link $L_{k}$ is blocked (shown in Fig.~\ref{altern_path}), $N_{k+1}$ will not receive expected data with high rate from $N_{k}$ and then replied NACK, and if $N_{k}$ receives nothing back or three consecutive NACK (through possible NLOS paths) from $N_{k+1}$ during its own transmitting slot $T_{s}$, both $N_{k}$ and $N_{k+1}$ will know the occurrence of blockage between them by the end of $T_{s}$. We denote this time as $t_{0}$.

\begin{figure}[htbp]
	\vspace{-0.2cm} 
	\setlength{\abovecaptionskip}{+0.1cm} 
	\centerline{\includegraphics[width=170mm]{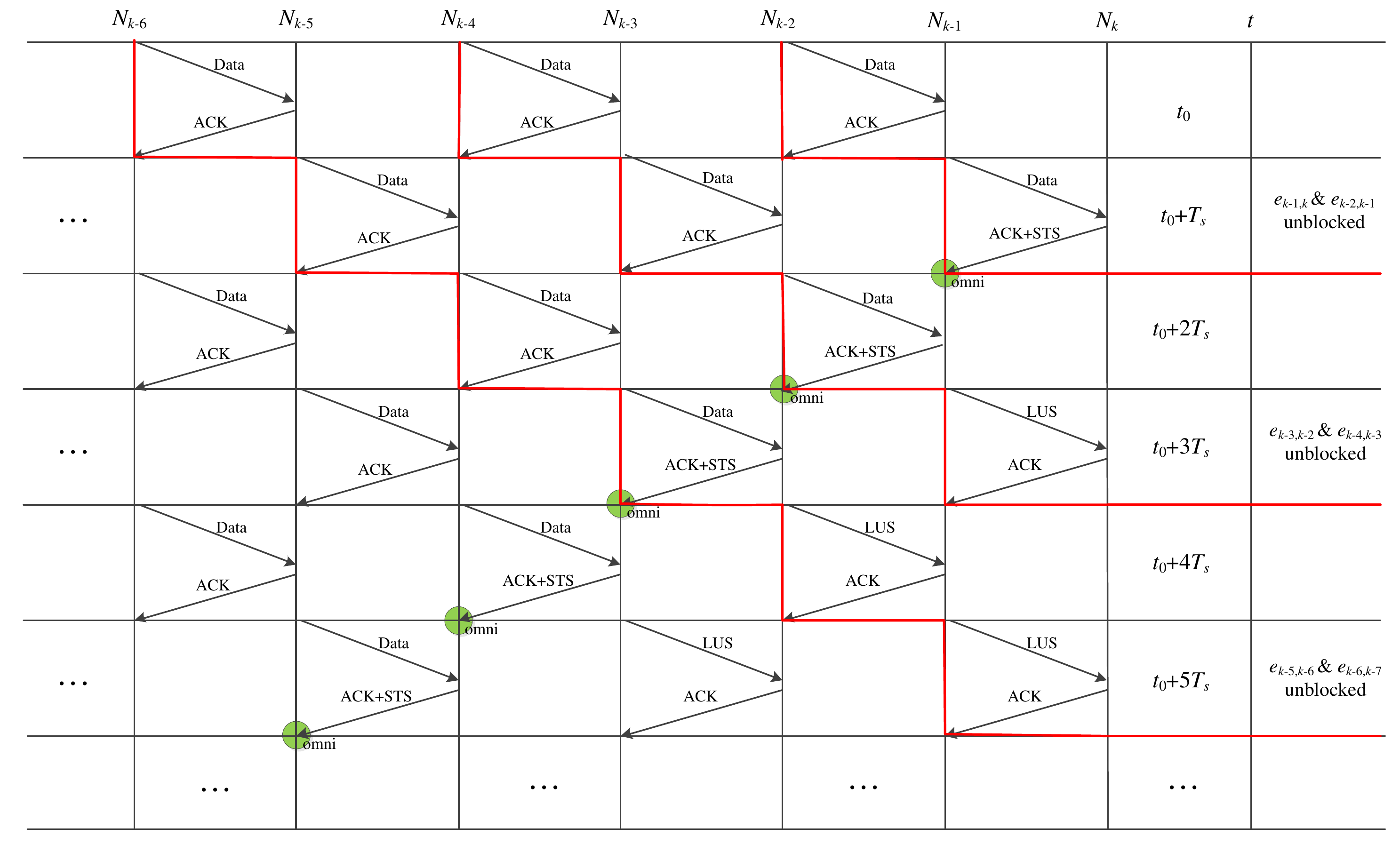}}
	\caption{Timing diagram of  $N_{k}$ and its left-side nodes.}
	\label{time-diag}
\end{figure}

\textbf{\textit{Step} 2.} Starting from $t_{0}$, $N_{k}$ waits for a period of time, which is called ``nodes preparation time" (NPT). Considering left-part and right-part nodes of $N_{k}$, NPT can be computed as follow:

\begin{equation}
 NPT = \left\{ \begin{array}{l}
\max \{ (\left\lceil {k/2} \right\rceil  \cdot 2 - 1),N - k - 2\}  \cdot {T_s},{\rm{~~~1}} \le k \le N{\rm{ - 2}} \\ 
(\left\lceil {k/2} \right\rceil  \cdot 2 - 1) \cdot {T_s},{\rm{~~~~~~~~~~~~~~~~~~~~~~~~~~}}k{\rm{ = }}N{\rm{ - 1, }}N \\ 
(N - 2) \cdot {T_s},{\rm{~~~~~~~~~~~~~~~~~~~~~~~~~~~~~~~~~~}}k{\rm{ = 0}} \\ 
\end{array} \right.
\end{equation}

\noindent where \textit{k} is the node index of $N_{k}$ and \textit{N} is the number of nodes in the topology. This NPT is consumed for two reasons:

\vspace{-0.3cm} 
\begin{enumerate}
	\item At the beginning of NPT, $N_{k+1}$ will send data as usual during its own transmitting time slot $T_{s}$, but appending a stop transmitting signal (STS) to $N_{k+2}$, then $N_{k+2}$ also transmits data and STS to next node in its time slot, therefore, after ($N$-$k$-$2$)$\cdot$$T_{s}$, all latter nodes of $N_{k+1}$, i.e. $N_{k+i}$ ($i$$\ge$2), will receive STS and know that one of previous original links is blocked, meanwhile switch their antennas to the omni-directional fashion from left to right once getting STS, which is convenient for receiving the following possible detection information. By the end of NPT, every node on the right side of $N_{k}$ completes the antenna switching process, and it could equip an additional omni-directional antenna, or adjust the beamwidth of its phase array antenna for this process.
	\item For the left side nodes of $N_{k}$, this NPT is not only used for avoiding BTD conflicts when multiple blockages occur concurrently, but also making these nodes stop transmitting data. As Fig.~\ref{time-diag} shows, after one $T_{s}$ from $t_{0}$, $N_{k}$ has received data from $N_{k-1}$, and it replied ACK appending STS, then $N_{k-1}$ would stop sending data to $N_{k}$ in next time slot and switch its antenna to omni-directional fashion for following possible detection. Meanwhile, $N_{k}$ could confirm that previous original links $e_{k-2,k-1}$ and $e_{k-1,k}$ are unblocked. After 2$T_{s}$, $N_{k-1}$ replied ACK appending STS after receiving data from $N_{k-2}$, then $N_{k-2}$ would stop sending data to $N_{k-1}$ in next time slot and switch its antenna to omni-directional fashion as well. After 3$T_{s}$, $N_{k}$ received the link-unblock sign (LUS) instead of data from $N_{k-1}$, so that it could know that $e_{k-4,k-3}$ and $e_{k-3,k-2}$ are also unobstructed. However, if $N_{k}$ did not receive data or received link-blocked sign (LBS) from $N_{k-1}$ by the end of $t_{0}+T_{s}$ or $t_{0}+3T_{s}$, that means another previous original link is blocked as well, so the previous node should first conduct the BTD process, and $N_{k}$ as the latter node will switch its antenna to the omni-directional fashion for possible coordination. If $N_{k}$ can still receive LUS from $N_{k-1}$ by the end of NPT, it will confirm that all previous original links are unobstructed, and it can continue the BTD process for this blockage on the link $L_{k}$. We can conclude previous original links information by Tab.~\ref{decision_NPT} from $N_{k}$'s perspective (``0" and ``1" denote the link is blocked or unblocked):
\end{enumerate}

\vspace{-0.4cm} 
\begin{table}[htbp]
	\centering
	\caption{~~Decision of previous original link from $N_{k}$ during NPT.}
	\vspace{-0.1cm} 
	\label{decision_NPT}
	\begin{tabular}{|p{2.0cm}|p{3.0cm}|p{3.0cm}|}	
		\hline
		\multicolumn{1}{|c|}{Time after $t_{0}$}	& \multicolumn{1}{c|}{Receive data/LUS from $N_{k-1}$} & \multicolumn{1}{c|}{Receive $1^{st}$ null/LBS from $N_{k-1}$}    \\
		\hline
		\multicolumn{1}{|c|}{$T_{s}$}	& \multicolumn{1}{c|}{$e_{k-1,k}$=1 \&\& $e_{k-2,k-1}$=1} & \multicolumn{1}{c|}{$e_{k-1,k}$=0 ~~$\parallel$ $e_{k-2,k-1}$=0}     \\
		\hline
		\multicolumn{1}{|c|}{3$T_{s}$}	& \multicolumn{1}{c|}{$e_{k-3,k-2}$=1 \&\& $e_{k-4,k-3}$=1} & \multicolumn{1}{c|}{$e_{k-3,k-2}$=0 $\parallel$ $e_{k-4,k-3}$=0}     \\
		\hline
				\multicolumn{1}{|c|}{5$T_{s}$}	& \multicolumn{1}{c|}{$e_{k-5,k-6}$=1 \&\& $e_{k-6,k-7}$=1} & \multicolumn{1}{c|}{$e_{k-5,k-6}$=0 $\parallel$ $e_{k-6,k-7}$=0}     \\
		\hline
				\multicolumn{1}{|c|}{...}	& \multicolumn{1}{c|}{...} & \multicolumn{1}{c|}{...}     \\
		\hline
				\multicolumn{1}{|c|}{($\left\lceil {k/2} \right\rceil$$\cdot$2-1)$\cdot$$T_{s}$}	& \multicolumn{1}{c|}{$e_{1,2}$=1 \&\& $e_{0,1}$=1} & \multicolumn{1}{c|}{$e_{1,2}$=0 $\parallel$ $e_{0,1}$=0}     \\
		\hline
	\end{tabular}
\end{table}

\vspace{-0.0cm} 

\begin{figure}[htbp]
	\vspace{-0.1cm} 
	\setlength{\abovecaptionskip}{+0.cm} 
	\centerline{\includegraphics[width=160mm]{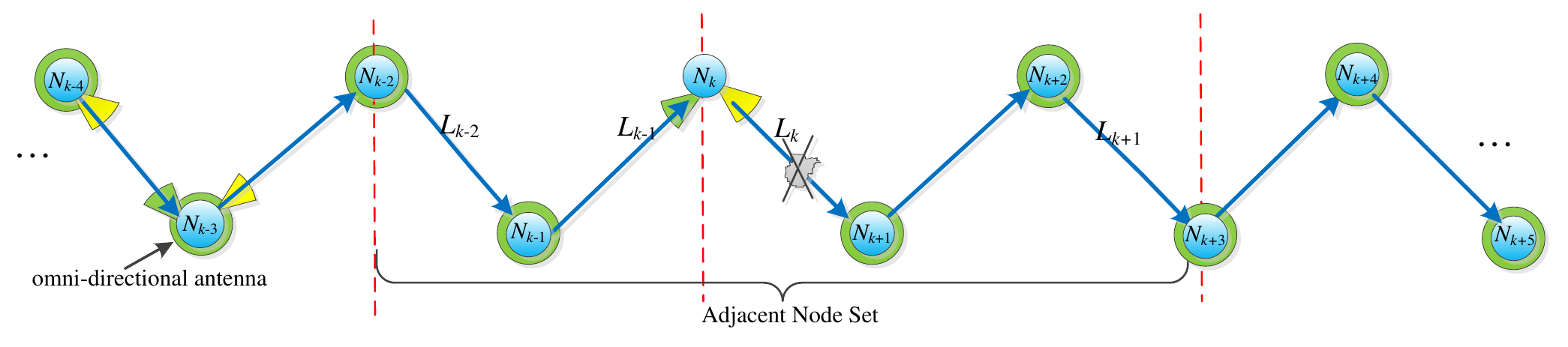}}
	\caption{The status of nodes after NPT.}
	\label{status_NPT}
\end{figure}

From above, $N_{k}$ will get the link information of all previous node pairs according the reception from $N_{k-1}$ during NPT, and by the end of NPT, all previous nodes of $N_{k}$ will stop transmitting data and switch to the omni-directional antenna for possible detection process. Fig.~\ref{status_NPT} shows the status of nodes in the topology after NPT.

\textbf{\textit{Step} 3.} After waiting for the time period NPT, $N_{k}$ starts to send blockage type detection preparation information (BTDP) to $N_{k-1}$ at $t_{0}$$'$ ($t_{0}$$'$=$t_{0}$+NPT), then $N_{k-1}$ will also forward this BTDP to $N_{k-2}$ after the time durance of transmitting BTDP along the original link $T_{s}$$'$ ($T_{s}$$'$$\ll$$T_{s}$), therefore, all left-side nodes of $N_{k}$ would receive BTDP after (\textit{k}-1)$\cdot$$T_{s}$$'$. The format of BTDP is shown in Tab.~\ref{BTDP_format}:

\begin{table}[htbp]
	\centering
	\caption{~~The format of BTDP.}
	\vspace{-0.1cm} 
	\label{BTDP_format}
	\begin{tabular}{|p{2.0cm}|p{3.0cm}|p{3.0cm}|p{3.0cm}|p{3.0cm}|}	
		\hline
		\multicolumn{1}{|c|}{NID}	& \multicolumn{1}{c|}{BLID} & \multicolumn{1}{c|}{STD}  & \multicolumn{1}{c|}{DTL} & \multicolumn{1}{c|}{DO}  \\
		\hline
     	\multicolumn{1}{|c|}{$N_{k}$}	& \multicolumn{1}{c|}{\textit{K}} & \multicolumn{1}{c|}{$t_{0}$$'$+2$\cdot$$T_{s}$$'$}  & \multicolumn{1}{c|}{$T_{k}$} & \multicolumn{1}{c|}{\{$N_{k}$:$t_{1}$; $N_{k-2}$:$t_{2}$; $N_{k-1}$:$t_{3}$; $N_{k+1}$:$t_{4}$; $N_{k+2}$:$t_{5}$; $N_{k+3}$:$t_{6}$\}, interval:$\bigtriangleup$}  \\  
		\hline
	\end{tabular}
\end{table}

\noindent where BID and BLID is the node ID and blocked link number, respectively, STD indicates start time of detection, DTL is the total length of this detection time, and DO represents the detection order of every participant appending respective starting time and interval. Before STD, the adjacent nodes $N_{k-1}$ and $N_{k-2}$ had received the BTDP and were ready for following detection process, and this is the end of the first phase.

\subsection{Detection Phase}
\begin{figure}[htbp]
	\vspace{-0.1cm} 
	\setlength{\abovecaptionskip}{+0.cm} 
	\centerline{\includegraphics[width=160mm]{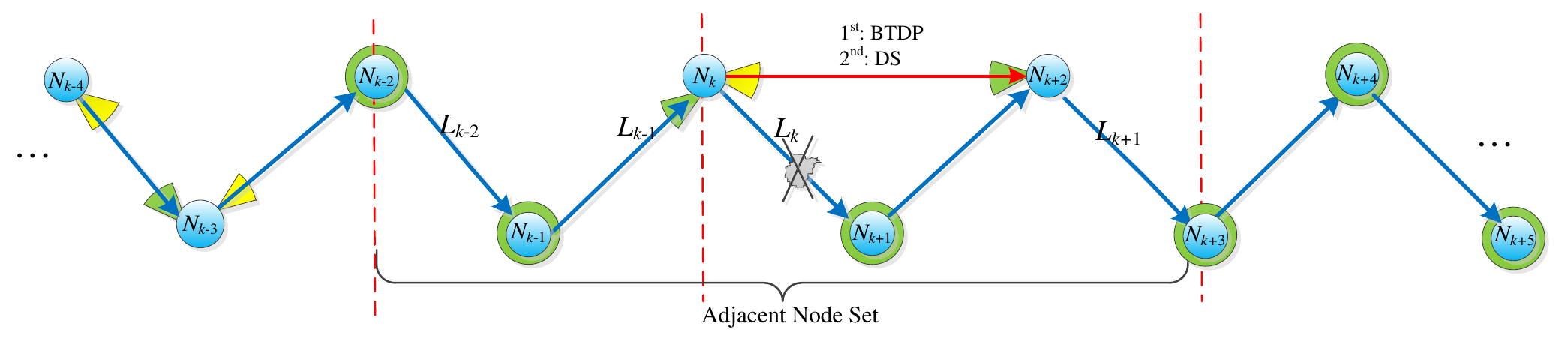}}
	\caption{The detection process between $N_{k}$ and $N_{k+2}$ from $t_{1}$ to $t_{2}$.}
	\label{detec_proc}
\end{figure}

Blockage type detection process starts at STD. Note that, before sending detection signal (DS), the node will adjust its directional antenna to align the target node's and send BTDP firstly, so that the possible end nodes of substituted paths (i.e. right-part nodes of $N_{k}$) can get the detection information and switch to the direction antenna aligning with the source node's, then reply an ACK. After receiving the ACK from the intended node, the DS will be sent by detector node for checking if this substituted path is available. After replying ACK for DS, every target node would recover the omni-directional antenna for next possible detection. In addition, during this phase, the block-link table \textit{A} is updated and shared by $N_{k}$ and its adjacent nodes, which includes all possible paths $e_{k+i,k+j}$ (-2$\le$$i$$\le$3, $i$+1$\le$$j$$\le$$i$+3) for the nodes in DO and corresponding values (``0"-not checked, ``1"-not blocked, ``inf"-blocked). The detailed steps are listed as follow:

\textbf{\textit{Step} 1}: At $t_{1}$, it is the turn of $N_{k}$ to check the available links in its $Sub_{k}$. $N_{k}$ first uses directional antenna to sends BTDP to $N_{k+2}$, and if receiving ACK from target node that has aligned the antenna to $N_{k}$ and is ready for detection, $N_{k}$ will send DS to check this substituted link $e_{k,k+2}$ (shown in Fig.~\ref{detec_proc}). When finishing detecting process with $N_{k+2}$, $N_{k}$ starts to checks the substituted link to $N_{k+3}$ in the same way. After checking every link in $Sub_{k}$, $N_{k}$ initializes the block-link table and sets the corresponding value for its checked links, then multicasts $A_{k}$ to adjacent nodes because each node in $Adj$ uses omni-directional antenna now. After that, $N_{k}$ switches its antenna to an omni-directional fashion for possible detection from other nodes.

\vspace{-0.1cm} 

\textbf{\textit{Step} 2}: At $t_{2}$ or $t_{3}$, it is detection time for left side nodes $N_{k-2}$ or $N_{k-1}$. Similar to $N_{k}$, $N_{k-2}$ or $N_{k-1}$ will firstly send BTDP to target node such as $N_{k+1}$, and then send DS if receiving ACK and the alignment of their antennas is done. After checking all the substituted paths from its \textit{Sub}, $N_{k-2}$ or $N_{k-1}$ updates and multicasts $A_{k}$, then switches to the omni-directional antenna.

\textbf{\textit{Step} 3}: From $t_{4}$ to $t_{6}$+$\bigtriangleup$ (i.e. the end of $N_{k+3}$'s detection time), the right-side nodes $N_{k+1}$, $N_{k+2}$ and $N_{k+3}$ will conduct detection process one by one in their respective time interval. After sending BTDP and receiving ACK, $N_{k+i}$ (1$\le$$i$$\le$3) starts to detect whether they can go to their latter hops, namely sending DS to check the available links from their $Tl$ (TX links set). If $N_{k+i}$ does not have any latter-hop links, it will reset all $e_{k+i-j,k+i}$ (1$\le$$i$$\le$3, 1$\le$$j$$\le$3) as $inf$ in $A_{k}$ for indicating that the possible alternative links into itself are actually not available, for example, if $N_{k+1}$ found that no available links for itself to go to next hops, it would update $e_{k-2,k+1}$ and $e_{k-1,k+1}$ as blocked status even if these two substituted links are available after checking by $N_{k-2}$ and $N_{k-1}$. After that, $N_{k+i}$ multicasts the new $A_{k}$ and switches antenna to omni-directional fashion.

\textbf{\textit{Step} 4}: After $t_{6}$+$\bigtriangleup$, $N_{k}$ can get the full block-link table $A_{k}$, which contains all link information of adjacent nodes, and the link that was not checked in $A_{k}$ would be viewed as blocked status due to the possible failed-node cases. Then $N_{k}$ can judge the type of blockage according to Tab. I in [6]. For example, if just $e_{k,k+3}$ or $e_{k-2,k+1}$ in $A_{k}$ is $inf$, the blockage is in \textit{Type} II for $N_{k}$ or $N_{k+1}$; But if just $e_{k-1,k+2}$ in $A_{k}$ is $inf$, it is the \textit{Type} III blockage; When all $e_{k,k+i}$ (2$\le$$i$$\le$3) or $e_{k-i,k+1}$ (1$\le$$i$$\le$2) are $inf$, the blockage is in \textit{Type} IV for $N_{k}$ or $N_{k+1}$; And it is the \textit{Type} I blockage when just $e_{k,k+1}$ is \textit{inf} in $A_{k}$. In other cases, the blockage can be always decomposed into these four types or their combination according to \textit{Theorem} I in [6].

Until now, the second phase of BTD is finished, and the next steps about relay path selection are described detailedly in [6] (Section III). After that, the new substituted path for blockage avoidance is generated by $N_{k}$, eg. selecting $e_{k,k+3}$ instead of $e_{k,k+1}$. Then $N_{k}$ will multicast Path Reconfiguration Message (PRM) to its adjacent nodes, which contains the new selected path (eg. $e_{k,k+3}$), and each of them that received PRM is required to multicast as well in case some adjacent nodes did not get this information due to blockage effect. Once the start and end node of this new substituted path receive PRM, they will use their directional antennas to align with each other. Particularly, if one of nodes in \textit{Adj} did not receive PRM until STD+DTL, that means no available paths can reach this failed node, and it does not need to transmit data but stays in present status until one of paths into itself is available. Note that, if the new substituted path (eg. $e_{k,k+2}$) breaks the original scheduling, i.e. respective transmitting time slot of nodes after this new path, the information about rescheduled time slots is also required to add in PRM and transmit to the nodes after this new substituted path.

\begin{figure}[htbp]
	\vspace{-0.1cm} 
	\setlength{\abovecaptionskip}{+0.cm} 
	\centerline{\includegraphics[width=160mm]{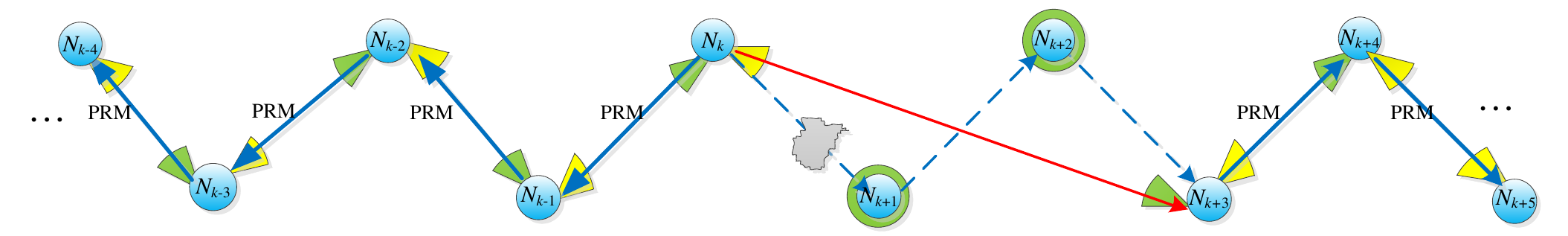}}
	\vspace{+0.15cm} 
	\caption{The process of transmitting PRM from STD+DTL to STT.}
	\label{trans_PRM}
\end{figure}


As Fig.~\ref{trans_PRM} shows, at the time \textit{t}=STD+DTL, the start and end nodes of the new path ($N_{k}$ and $N_{k+3}$) will start to transmit PRM to previous node and latter node, respectively. Once receiving PRM, each node in the topology forwards PRM to previous or latter node as well, and switches its antenna to the directional fashion and aligns the TX/RX antenna to its latter/previous node. When the source node receives PRM, it starts transmitting data, and the latter nodes will recover data transmission process once receiving data from previous nodes.

In summary, Algorithm 1 (a)-(c) in Appendix shows the pseudocode for the whole process from $t_{1}$ to the time that data transmission has been recovered, which contains respective works of the leader node $N_{k}$ and its adjacent nodes $N_{k-m}$ (1$\le$$m$$\le$2), $N_{k+n}$ (1$\le$$n$$\le$3) during this process.

\vspace{+0.5cm}
\noindent \textbf{\large References}
\vspace{+0.2cm}

\noindent [1] J. Du, E. Onaran, et al., ``Gbps user rates using mmWave relayed backhaul with high gain antennas", \textit{IEEE Journal on Selected Areas in Communications}, 2017: 1363-1372.

\vspace{+0.1cm}
\noindent [2] Q. Hu and D. Blough, ``Optimizing millimeter-wave backhaul networks in roadside environments", \textit{Proc. of IEEE International Conference on Communications}, 2018.

\vspace{+0.1cm}
\noindent [3] ``Terragraph: Solving the urban bandwidth challenge", Facebook Connectivity, 2019.

\vspace{+0.1cm}
\noindent [4] A. Dimas, D. Kalogerias, and A. Petropulu, ``Cooperative beamforming with predictive relay selection for urban mmWave communications", \textit{IEEE Access} (2019): 157057-157071.

\vspace{+0.1cm}
\noindent [5] Y. Liu and D. Blough, ``Analysis of Blockage Effects on Roadside Relay-assisted mmWave Backhaul Networks", \textit{Proc. of IEEE International Conference on Communications}, 2019.

\vspace{+0.1cm}
\noindent [6] Y. Liu, Q. Hu, and D. Blough, ``Blockage Avoidance in Relay Paths for Roadside mmWave Backhaul Networks", \textit{Proc. of IEEE Symposium on Personal, Indoor, and Mobile Radio Communications}, 2018.

\vspace{+0.1cm}
\noindent [7] Y. Liu, Q. Hu and D. Blough, ``Joint Link-level and Network-level Reconfiguration for mmWave Backhaul Survivability in Urban Environments", \textit{Proc. of Modeling, Analysis and Simulation of Wireless and Mobile Systems}, ACM, 2019.

\vspace{+0.1cm}
\noindent [8] Y. Liu, Q. Hu and D. Blough, ``Joint Link-level and Network-level Reconfiguration for Urban mmWave Wireless Backhaul Networks", \textit{Computer Communications}, 2020.

\vspace{+0.1cm}
\noindent [9] T. Bai, R. Vaze, H. Robert, ``Analysis of blockage effects on urban cellular networks", \textit{IEEE Transactions on Wireless Comm.}, 2014.

\vspace{+0.1cm}
\noindent [10] T. Bai, H. Robert, ``Coverage and rate analysis for millimeter-wave cellular networks", \textit{IEEE Transactions on Wireless Comm.}, 2015

\newpage
\begin{center}
	\Large \textbf{\textbf{Appendix}}
\end{center}

\vspace{-0.2cm} 
\begin{algorithm}[h!]
	\caption{(a) The status of the leader node ($N_{k}$) for blockage detection process}
	\begin{algorithmic}[1]
     \STATE ~~~~~~~~~~~~\textbf{while}($t_{1}$$\le$$t$$<$$t_{2}$) \textbf{do}
     \STATE ~~~~~~~~~~~~~~~~~~\textbf{for} \textit{link} with \textit{end node i} in $Sub_{k}$ \textbf{do}
     \STATE ~~~~~~~~~~~~~~~~~~~~~~~AlignAntenna(\textit{i}); // align the antenna to node \textit{i}
     \STATE ~~~~~~~~~~~~~~~~~~~~~~~BTDPsend(\textit{i}); // send BTDP to node \textit{i}
     \STATE ~~~~~~~~~~~~~~~~~~~~~~~\textbf{if} (ACK) \textbf{do} //receive an ACK
     \STATE ~~~~~~~~~~~~~~~~~~~~~~~~~~~DSsend(\textit{\textit{i}}); // send DS to node \textit{i}
     \STATE ~~~~~~~~~~~~~~~~~~~~~~~~~~~$A_{k}$.\textit{init}; // initialize $A_{k}$
     \STATE ~~~~~~~~~~~~~~~~~~~~~~~~~~~\textbf{if}(ACK) \textbf{do}
     \STATE ~~~~~~~~~~~~~~~~~~~~~~~~~~~~~~$A_{k}.set(e_{k,i})$ = 1; // update $e_{k,i}$=1 in $A_{k}$
     \STATE ~~~~~~~~~~~~~~~~~~~~~~~~~~~\textbf{else}
     \STATE ~~~~~~~~~~~~~~~~~~~~~~~~~~~~~~$A_{k}.set(e_{k,i})$ = 0;
     \STATE ~~~~~~~~~~~~~~~~~~~~~~~~~~~\textbf{endif}
     \STATE ~~~~~~~~~~~~~~~~~~~~~~~\textbf{else}
     \STATE ~~~~~~~~~~~~~~~~~~~~~~~~~~~$A_{k}.set(e_{k,i})$ = 0;
     \STATE ~~~~~~~~~~~~~~~~~~~~~~~\textbf{endif}
     \STATE ~~~~~~~~~~~~~~~~~~\textbf{endfor}
     \STATE ~~~~~~~~~~~~~~~~~~mCast($A_{k}$); //multicast $A_{k}$ to its adjacent nodes
     \STATE ~~~~~~~~~~~~~~~~~~OmniAntenna($N_{k}$); // Switch to omni-directional antenna
     \STATE ~~~~~~~~~~~~~~~~~~\textbf{wait}; // as a listener
     \STATE ~~~~~~~~~~~~\textbf{endwhile}
     \STATE ~~~~~~~~~~~~\textbf{while}($t_{2}$$\le$$t$$<$$t_{6}$+$\bigtriangleup$) \textbf{do}   // as a listener
     \STATE ~~~~~~~~~~~~~~~~~~\textbf{if}(\textit{recv} BTDP $\parallel$ DS from node \textit{j}) \textbf{do}
     \STATE ~~~~~~~~~~~~~~~~~~~~AlignAntenna(\textit{j});
     \STATE ~~~~~~~~~~~~~~~~~~~~ACK(\textit{j});
     \STATE ~~~~~~~~~~~~~~~~~~\textbf{endif}
     \STATE ~~~~~~~~~~~~~~~~~~\textbf{if} (\textit{recv} $A_{k}$)
     \STATE ~~~~~~~~~~~~~~~~~~~~~$A_{k}.update(A_{k})$; // replace old $A_{k}$ with received $A_{k}$
     \STATE ~~~~~~~~~~~~~~~~~~\textbf{endif}
     \STATE ~~~~~~~~~~~~\textbf{endwhile}
     \STATE ~~~~~~~~~~~~\textbf{while}($t_{6}$+$\bigtriangleup$$\le$$t$$<$STD+DTL) \textbf{do}  // as a blockage type detector
     \STATE ~~~~~~~~~~~~~~~~~~\textit{Type}=BTD($A_{k}$) // Detect Blockage Type
     \STATE ~~~~~~~~~~~~~~~~~~Reconf(\textit{Type}); // Reconfigure new path by HTPR algorithm in [6])
     \STATE ~~~~~~~~~~~~~~~~~~mCast(PRM);
     \STATE ~~~~~~~~~~~~~~~~~~\textbf{if} (\textit{NodeIndex} in PRM) \textbf{do} // Prepare to transmission
     \STATE ~~~~~~~~~~~~~~~~~~~~~~DirectAntenna($N_{k}$); // Switch to directional antenna
     \STATE ~~~~~~~~~~~~~~~~~~~~~~AlignAntenna(\textit{PreNode} and \textit{LatterNode} in PRM);
     \STATE ~~~~~~~~~~~~~~~~~~\textbf{endif}
     \STATE ~~~~~~~~~~~~~~~~~~\textbf{wait}; // idle time if process has been finished
     \STATE ~~~~~~~~~~~~\textbf{endwhile}
	\end{algorithmic}
\end{algorithm}

\begin{algorithm}[h!]
\setcounter{algorithm}{0}
	\caption{(b) The status of $N_{k-m}$ (1$\le$$m$$\le$2) for blockage detection process}
	\begin{algorithmic}[1]
		\STATE ~~~~~~~~~~~~\textbf{while}($t_{4-m}$$\le$$t$$<$$t_{5-m}$) \textbf{do}
		\STATE ~~~~~~~~~~~~~~~~~~\textbf{for} \textit{link} with \textit{end node i} in $Sub_{k-m}$ \textbf{do}
		\STATE ~~~~~~~~~~~~~~~~~~~~~~~AlignAntenna(\textit{i}); // align the antenna to node \textit{i}
		\STATE ~~~~~~~~~~~~~~~~~~~~~~~BTDPsend(\textit{i}); // send BTDP to node \textit{i}
		\STATE ~~~~~~~~~~~~~~~~~~~~~~~\textbf{if} (ACK) \textbf{do} //receive an ACK
		\STATE ~~~~~~~~~~~~~~~~~~~~~~~~~~~DSsend(\textit{\textit{i}}); // send DS to node \textit{i}
		\STATE ~~~~~~~~~~~~~~~~~~~~~~~~~~~\textbf{if}(ACK) \textbf{do}
		\STATE ~~~~~~~~~~~~~~~~~~~~~~~~~~~~~~$A_{k}.set(e_{k,i})$ = 1; // update $e_{k,i}$=1 in $A_{k}$
		\STATE ~~~~~~~~~~~~~~~~~~~~~~~~~~~\textbf{else}
		\STATE ~~~~~~~~~~~~~~~~~~~~~~~~~~~~~~$A_{k}.set(e_{k,i})$ = 0;
		\STATE ~~~~~~~~~~~~~~~~~~~~~~~~~~~\textbf{endif}
		\STATE ~~~~~~~~~~~~~~~~~~~~~~~\textbf{else}
		\STATE ~~~~~~~~~~~~~~~~~~~~~~~~~~~$A_{k}.set(e_{k,i})$ = 0;
		\STATE ~~~~~~~~~~~~~~~~~~~~~~~\textbf{endif}
		\STATE ~~~~~~~~~~~~~~~~~~\textbf{endfor}
		\STATE ~~~~~~~~~~~~~~~~~~mCast($A_{k}$); //multicast $A_{k}$ to its adjacent nodes
		\STATE ~~~~~~~~~~~~~~~~~~OmniAntenna($N_{k-m}$); // Switch to omni-directional antenna
		\STATE ~~~~~~~~~~~~~~~~~~\textbf{wait}; // as a listener
		\STATE ~~~~~~~~~~~~\textbf{endwhile}
		\STATE ~~~~~~~~~~~~\textbf{while}($t_{6}$+$\bigtriangleup$$\le$$t$$<$STD+DTL) \textbf{do}  // as a broadcaster
		\STATE ~~~~~~~~~~~~~~~~~~\textbf{if} (\textit{recv} PRM) \textbf{do}
		\STATE ~~~~~~~~~~~~~~~~~~~~~mCast(PRM);
		\STATE ~~~~~~~~~~~~~~~~~~~~~\textbf{if} (\textit{k-m} in PRM) \textbf{do} // Prepare to transmission
		\STATE ~~~~~~~~~~~~~~~~~~~~~~~~~DirectAntenna($N_{k-m}$); // Switch to directional antenna
		\STATE ~~~~~~~~~~~~~~~~~~~~~~~~~AlignAntenna(\textit{PreNode} and \textit{LatterNode} in PRM);
		\STATE ~~~~~~~~~~~~~~~~~~~~~\textbf{endif}
	    \STATE ~~~~~~~~~~~~~~~~~~\textbf{endif}
		\STATE ~~~~~~~~~~~~~~~~~~\textbf{wait}; // idle time if process has been finished
		\STATE ~~~~~~~~~~~~\textbf{endwhile}
		\STATE ~~~~~~~~~~~~\textbf{while}(other time) \textbf{do}
		\STATE ~~~~~~~~~~~~~~~~~~\textbf{if} (\textit{recv} $A_{k}$) \textbf{do}
		\STATE ~~~~~~~~~~~~~~~~~~~~~$A_{k}.update(A_{k})$;
		\STATE ~~~~~~~~~~~~~~~~~~\textbf{endif}
		\STATE ~~~~~~~~~~~~\textbf{endwhile}
		
	\end{algorithmic}
\end{algorithm}

\begin{algorithm}[h!]
\setcounter{algorithm}{0}
	\caption{(c) The status of $N_{k+n}$ (1$\le$$n$$\le$3) for blockage detection process}
	\begin{algorithmic}[1]
		\STATE ~~~~~~~~~~~~\textbf{while}($t_{n+3}$$\le$$t$$<$$t_{n+4}$) \textbf{do}
		\STATE ~~~~~~~~~~~~~~~~~~\textbf{for} \textit{link} with \textit{end node i} in $Tl_{k+n}$ \textbf{do}  ~~// check all TX links
		\STATE ~~~~~~~~~~~~~~~~~~~~~~~AlignAntenna(\textit{i}); // align the antenna to node \textit{i}
		\STATE ~~~~~~~~~~~~~~~~~~~~~~~BTDPsend(\textit{i}); // send BTDP to node \textit{i}
		\STATE ~~~~~~~~~~~~~~~~~~~~~~~\textbf{if} (ACK) \textbf{do} //receive an ACK
		\STATE ~~~~~~~~~~~~~~~~~~~~~~~~~~~DSsend(\textit{\textit{i}}); // send DS to node \textit{i}
		\STATE ~~~~~~~~~~~~~~~~~~~~~~~~~~~\textbf{if}(ACK) \textbf{do}
		\STATE ~~~~~~~~~~~~~~~~~~~~~~~~~~~~~~$A_{k}.set(e_{k,i})$ = 1; // update $e_{k,i}$=1 in $A_{k}$
		\STATE ~~~~~~~~~~~~~~~~~~~~~~~~~~~\textbf{else}
		\STATE ~~~~~~~~~~~~~~~~~~~~~~~~~~~~~~$A_{k}.set(e_{k,i})$ = 0;
		\STATE ~~~~~~~~~~~~~~~~~~~~~~~~~~~\textbf{endif}
		\STATE ~~~~~~~~~~~~~~~~~~~~~~~\textbf{else}
		\STATE ~~~~~~~~~~~~~~~~~~~~~~~~~~~$A_{k}.set(e_{k,i})$ = 0;
		\STATE ~~~~~~~~~~~~~~~~~~~~~~~\textbf{endif}
		\STATE ~~~~~~~~~~~~~~~~~~\textbf{endfor}
		\STATE ~~~~~~~~~~~~~~~~~~\textbf{if} (all links in $Tl_{k+n}$ equal to zero) \textbf{do} // reset values of unavailable links in $A_{k}$
		\STATE ~~~~~~~~~~~~~~~~~~~~~\textbf{for} \textit{j}=1 to 3 \textbf{do}
		\STATE ~~~~~~~~~~~~~~~~~~~~~~~~~~$A_{k}.set(e_{k+n-j,k+n})$ = 0;
		\STATE ~~~~~~~~~~~~~~~~~~~~~\textbf{endfor}
		\STATE ~~~~~~~~~~~~~~~~~~\textbf{endif}
		
		\STATE ~~~~~~~~~~~~~~~~~~mCast($A_{k}$); //multicast $A_{k}$ to its adjacent nodes
		\STATE ~~~~~~~~~~~~~~~~~~OmniAntenna($N_{k+n}$); // Switch to omni-directional antenna
		\STATE ~~~~~~~~~~~~~~~~~~\textbf{wait}; // as a listener
		\STATE ~~~~~~~~~~~~\textbf{endwhile}
		\STATE ~~~~~~~~~~~~\textbf{while}($t_{1}$$\le$$t$$<$$t_{k+n}$ $\parallel$ $t_{k+n+1}$$\le$$t$$<$$t_{6}$+$\bigtriangleup$) // as a listener
		\STATE ~~~~~~~~~~~~~~~~~~\textbf{if} (\textit{recv} BTDP $\parallel$ DS from node \textit{a}) \textbf{do}
		\STATE ~~~~~~~~~~~~~~~~~~~~~~AlignAntenna(\textit{a});
		\STATE ~~~~~~~~~~~~~~~~~~~~~~ACK(\textit{a});
		\STATE ~~~~~~~~~~~~~~~~~~\textbf{endif}
		\STATE ~~~~~~~~~~~~~~~~~~\textbf{if} (\textit{recv} $A_{k}$) \textbf{do}
		\STATE ~~~~~~~~~~~~~~~~~~~~~$A_{k}.update(A_{k})$;
		\STATE ~~~~~~~~~~~~~~~~~~\textbf{endif}
		\STATE ~~~~~~~~~~~~\textbf{endwhile}
		\STATE ~~~~~~~~~~~~\textbf{while}($t_{6}$+$\bigtriangleup$$\le$$t$$<$STD+DTL) \textbf{do}  // as a broadcaster
		\STATE ~~~~~~~~~~~~~~~~~~\textbf{if} (\textit{recv} PRM) \textbf{do}
		\STATE ~~~~~~~~~~~~~~~~~~~~~mCast(PRM);
		\STATE ~~~~~~~~~~~~~~~~~~~~~\textbf{if} (\textit{k}+\textit{n} in PRM) \textbf{do} // Prepare to transmission
		\STATE ~~~~~~~~~~~~~~~~~~~~~~~~~DirectAntenna($N_{k+n}$); // Switch to directional antenna
		\STATE ~~~~~~~~~~~~~~~~~~~~~~~~~AlignAntenna(\textit{PreNode} and \textit{LatterNode} in PRM);
		\STATE ~~~~~~~~~~~~~~~~~~~~~\textbf{endif}
		\STATE ~~~~~~~~~~~~~~~~~~\textbf{endif}
		\STATE ~~~~~~~~~~~~~~~~~~\textbf{wait}; // idle time if process has been finished
		\STATE ~~~~~~~~~~~~\textbf{endwhile}
		
	\end{algorithmic}
\end{algorithm}

\end{document}